# The 23andMe Data Breach: Analyzing Credential Stuffing Attacks, Security Vulnerabilities, and Mitigation Strategies

Ryan Holthouse, Serena Owens, Suman Bhunia
*Department of Computer Science and Software Engineering, Miami University, Oxford, Ohio, USA 45056*
Email: holthorm@miamioh.edu, owenssj3@miamioh.edu, bhunias@miamioh.edu

*Abstract*—In October 2023, 23andMe, a prominent provider of personal genetic testing, ancestry, and health information services, suffered a significant data breach orchestrated by a cybercriminal known as "Golem." Initially, approximately 14,000 user accounts were compromised by a credential smear attack, exploiting reused usernames and passwords from previous data leaks. However, due to the interconnected nature of 23andMe's DNA Relatives and Family Tree features, the breach expanded exponentially, exposing sensitive personal and genetic data of approximately 5.5 million users and 1.4 million additional profiles. The attack highlights the increasing threat of credential stuffing, exacerbated by poor password hygiene and the absence of robust security measures such as multi-factor authentication (MFA) and rate limiting. In response, 23andMe mandated password resets, implemented email-based two-step verification, and advised users to update passwords across other services. This paper critically analyzes the attack methodology, its impact on users and the company, and explores potential mitigation strategies, including enhanced authentication protocols, proactive breach detection, and improved cybersecurity practices. The findings underscore the necessity of stronger user authentication measures and corporate responsibility in safeguarding sensitive genetic and personal data.

## 1. Introduction

23andMe is a company based in San Fransisco, California, that provides personal genetic testing. They use genotyping to analyze your DNA. This means they look at specific locations in your genome that are known to differ between groups of people. They then turn those results into personalized genetic reports on everything from ancestry composition to traits to genetic health risks [1]. 23andMe is best known for its health reports (an optional purchase) and enormous testing pool, second only to AncestryDNA. Founded in April of 2006, 23andMe became the first company to begin offering autosomal DNA testing for ancestry, pioneering the saliva-based testing medium that is still used today [2]. In May 2023, 23andMe reported their customer base to be over 14 million users. These incidents demonstrate the urgent need for improved password hygiene and multi-factor authentication.Fig. 1 draws a timeline of the 23AndMe data breach.

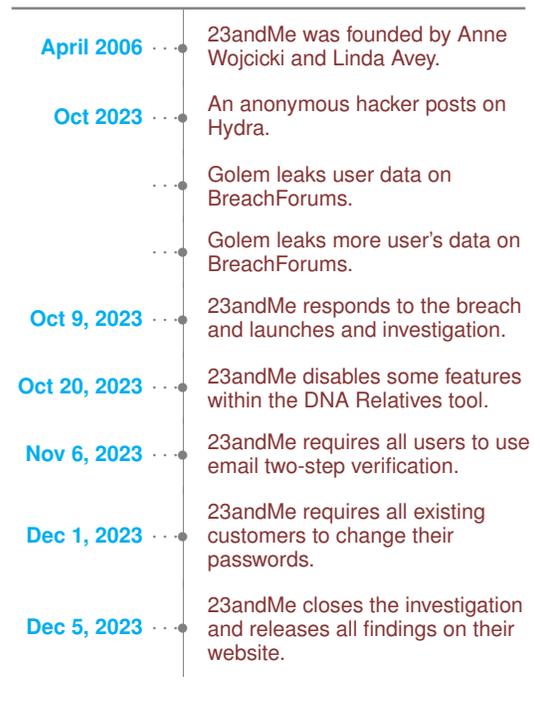

Figure 1: Timeline of 23AndMe Data Breach.

In early October 2023, 23andMe fell victim to a data breach caused by a cybercriminal known as Golem. Golem claimed to have leaked over 4 million users' data on a hacker forum called BreachForums [3]. As proof of the breach, Golem published the alleged data of one million users of Jewish Ashkenazi descent and 100,000 Chinese users, asking would-be buyers for $1 to $10 for the data per individual account. The data includes personal information such as display name, sex, birth year, and some details about genetic ancestry results. It may also include some more specific geographic ancestry information. The information did not appear to include actual, raw genetic data.

Two weeks later, Golem advertised the alleged records of another four million people on the same hacking forum. They claimed the dataset contained information on people who come from Great Britain, including data from "the wealthiest people living in the U.S. and Western Europe on this list"; however, this statement has not yet been confirmed to be true [4]. Before Golem's advertisement on BreachForums, another hacker took to a different forum, Hydra, where they also advertised a set of 23andMe user data. The hacker claimed to have 300 terabytes of 23andMe user data, though the hacker did not provide any evidence for this claim [4]. According to a Tech Crunch analysis, the data from this data breach matched some genealogical records published online by hobbyists and genealogists. The two sets of information were formatted differently but contained some of the same unique user and generic data, suggesting the data leaked by the hacker was at least in part authentic 23andMe customer data [5].

Golem's method for accessing this data is credential stuffing. Credential stuffing is the automated injection of stolen usernames and passwords into website login forms to fraudulently gain access to user accounts [6]. Since many people tend to reuse usernames and passwords, many people have their credentials and accounts stolen through previous data breaches that disclose this information [6]. As more information is being leaked, knowing the dangers of credential stuffing and creating strong and unique passwords across all websites is even more important. Some similar data breaches to this one include the Linkedin Users Data Breach, the Paypal Data Breach, and Norton Lifelock Data Breach [3].

In a press release from 23andMe, they noted that Golem was able to access less than 0.1%, or roughly 14,000 user accounts, of the existing 14 million 23andMe customers through credential stuffing. The hacker used the compromised credential-stuffed accounts to access the information included in a significant number of DNA Relatives profiles (approximately 5.5 million) and Family Tree feature profiles (approximately 1.4 million), each of which were connected to the compromised accounts, allowing them to grow their reach of compromised data. In response to the breach, a spokesperson from 23andMe confirmed the authenticity of the data found on the hacker forums. The company has been open about the nature of the attack, that the attackers utilized credentials exposed in other data breaches to infiltrate 23andMe accounts and gather sensitive data. Since detecting the breach, 23andMe emailed all customers to notify them of the investigation. They also required every 23andMe customer to reset their password. In addition, 23andMe now requires all new and existing customers to login using two-step verification. Along with these notifications, 23andMe added a page on their website called "Addressing Data Security Concerns," which detailed all aspects of their investigation along with their findings.

This paper provides an in-depth analysis of the 23andMe data breach, examining the company's and attackers' backgrounds, the specific methodologies used to compromise user data, and the broader implications for cybersecurity and data privacy. Furthermore, it evaluates the attack's impact on both the company and its customers while exploring potential mitigation strategies. These solutions include best practices for individual users as well as policy changes that companies like 23andMe can implement to enhance their security posture.

The remainder of this paper is structured as follows: Section 2 provides background on 23andMe's services and the cybercriminal behind the breach, highlighting why genetic testing platforms are prime targets for cyberattacks. Section 3 examines the attack methodology, detailing the credential stuffing technique and how the DNA Relatives and Family Tree features contributed to the large-scale exposure of user data. Section 4 explores the impact of the breach, including its consequences for affected users, potential legal and regulatory implications, and broader cybersecurity concerns. Section 5 discusses defense solutions, evaluating strategies such as multi-factor authentication, rate limiting, passwordless authentication, and user awareness programs to mitigate future attacks. Finally, Section 6 concludes the study by summarizing key findings and offering recommendations for enhancing cybersecurity measures in online genetic testing platforms.

## 2. Background

The 23andMe data breach represents a significant cybersecurity incident in the realm of personal genetic testing. Given the sensitive nature of genetic information, any compromise of such data raises serious privacy, ethical, and security concerns. This section provides an overview of 23andMe's services and the features that contributed to the breach, followed by an analysis of the attacker, Golem, their methods, motivations, and impact on data security. Understanding both the company's vulnerabilities and the attacker's approach is crucial to assessing the broader implications of this breach and the potential risks faced by users and similar online platforms.

### 2.1. 23andMe and Its Vulnerabilities

23andMe is a leading genetic testing company that offers a range of services, including ancestry analysis, health risk assessments, and genealogical connections. As a platform handling highly sensitive personal and genetic data, it presents an attractive target for cybercriminals. Among its features, DNA Relatives and Family Tree tools stand out as particularly data-rich environments where users voluntarily share genetic information with potential relatives.

The DNA Relatives feature is one of the most interactive components of 23andMe, designed to help users identify and connect with genetic relatives [7]. It works by comparing a user's DNA with that of others who have opted into the feature, identifying matching segments to predict relationships. Longer and more numerous shared segments typically indicate closer ancestry. This tool allows users to see shared genetic markers, estimated relationships, and other personal details, which are disclosed based on individual privacy settings. However, this feature also introduces significant security risks: if an attacker gains access to even a small number of user accounts, they can potentially retrieve data from thousands of interconnected profiles. In the 23andMe breach, the attackers leveraged this networked structure to expand their reach beyond the initially compromised accounts, ultimately exposing data from over 5.5 million DNA Relatives profiles and 1.4 million Family Tree profiles.

With a customer base exceeding 14 million users, many of whom choose to share personal data within the platform, 23andMe became a high-value target for cybercriminals. The interconnected nature of its services allowed attackers to exponentially scale the breach beyond the initial point of entry.

### 2.2. The Attacker: Golem

The individual behind the 23andMe data breach operates under the alias "Golem" and is a known figure in underground cybercrime forums. Golem joined BreachForums in September 2023 [8], and as of October 19, 2023, had posted seven times, participated in a single discussion thread, and held a "God User" reputation score of 35. Achieving this rank required spending at least £50 in cryptocurrency, suggesting a level of dedication to illicit activities beyond casual hacking.

Golem first announced the 23andMe breach by posting stolen genetic datasets for sale on BreachForums, claiming to have millions of user records. One of the most controversial aspects of the breach was Golem's targeting of specific demographic groups, including one million users of Jewish Ashkenazi descent and 100,000 Chinese users. Later, Golem expanded their claims, stating that they possessed data on some of the wealthiest individuals from the U.S. and Western Europe, although this assertion remains unverified [4].

In one of their posts, Golem attempted to justify their actions, writing: "After all, there are innocent people in these data. They don't need to be afraid; your important data is safer in [these] hands than with 23andMe."

Golem also invited potential buyers to contact them directly for access to more detailed or raw data, further indicating their intent to monetize stolen genetic and personal information.

Golem's Motivations and Risks Golem's reluctance to release all data immediately and focus on elite individuals led to speculation about their motives. Some observers likened their approach to a modern-day Robin Hood, exposing the data of the wealthy while sparing others. However, unlike a whistleblower or hacktivist, Golem did not use the data for advocacy or ethical concerns but instead commercialized it, making it available to the highest bidder.

Regardless of their claims or selective disclosures, Golem's actions constituted a severe breach of privacy. By exploiting credential stuffing techniques and the interconnectivity of 23andMe's features, they were able to access and distribute sensitive personal data at an unprecedented scale.

Figure 2 displays Golem's post on BreachForums, showcasing their willingness to share stolen data with interested buyers. While little is known about their identity beyond this attack, their behavior, access to high-profile data, and monetization strategies suggest they are a highly dangerous threat actor in the cybercrime landscape.

### 3. Attack Methodology

Cyberattacks are continuously evolving, with adversaries leveraging a diverse range of methodologies to infiltrate systems. In the case of the 23andMe data breach, the techniques used were relatively unsophisticated yet highly effective, centering around brute-force attacks and credential stuffing. The breach exploited recycled user credentials, highlighting the vulnerabilities associated with password reuse and weak authentication mechanisms.

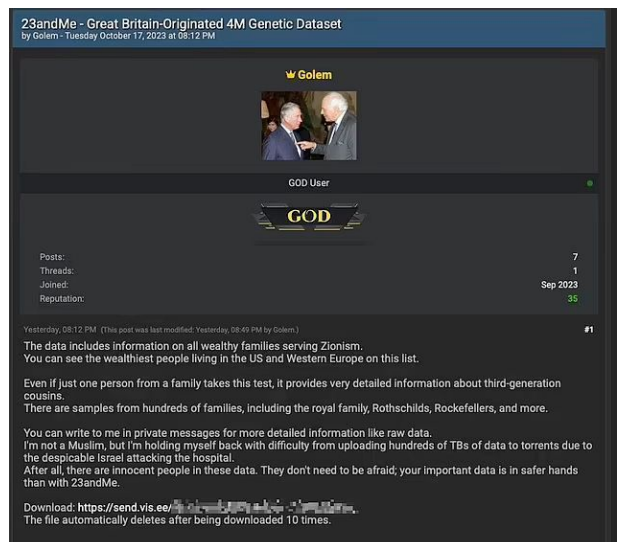

Figure 2: Golem's post on BreachForums

### 3.1. Brute Force and Credential Stuffing Attacks

Brute-force attacks are among the most fundamental cybersecurity threats, relying on automated scripts to systematically attempt passwords until the correct one is found. This method is inherently inefficient, as it depends on extensive computational resources and time. However, when paired with credential stuffing—a technique where attackers use previously leaked usernames and passwords from past data breaches—the probability of success increases substantially [9].

Research suggests that brute-force attacks can be mitigated through simple security measures, such as limiting the number of login attempts within a specific timeframe. Florêncio et al. demonstrated that enforcing a 24-hour lockout after three incorrect login attempts could reduce the probability of successfully brute-forcing a six-digit PIN to less than 1% over a ten-year period [10]. However, these estimations assume purely random guessing. In reality, attackers leverage data from previous breaches and human tendencies in password creation, making their attacks significantly more efficient.

### 3.2. Common Password Vulnerabilities

Weak password practices remain a persistent security flaw, as users frequently reuse simple and easily guessed passwords across multiple platforms. Studies analyzing leaked password datasets have consistently revealed a preference for weak, predictable credentials. Yue Li et al. examined the 12306 dataset, which contained more than 300,000 compromised Chinese passwords, and found that the most common password was 123456, which represents 0. 296% of the dataset, equivalent to nearly 900 accounts that could be easily compromised [11]. Table 1

Table 1: Table of the most frequent passwords held within the 12306 dataset, recreated from [11].

| Rank | Password | Amount | Percentage |
|---|---|---|---|
| 1 | 123456 | 389 | 0.296% |
| 2 | a123456 | 280 | 0.213% |
| 3 | 123456a | 165 | 0.125% |
| 4 | 5201314 | 160 | 0.121% |
| 5 | 111111 | 156 | 0.118% |
| 6 | woani1314 | 134 | 0.101% |
| 7 | qq123456 | 98 | 0.074% |
| 8 | 123123 | 97 | 0.073% |
| 9 | 000000 | 96 | 0.073% |
| 10 | 1qaz2wsx | 92 | 0.070% |

provides the most frequent passwords held within the 12306 dataset.

Credential stuffing attacks exploit this widespread password reuse, enabling attackers to breach multiple accounts using the same set of leaked credentials. This technique has been instrumental in numerous high-profile cyber incidents, including the Canva data breach [12] and the COMB database leak, which exposed over three billion credentials [13]. A study by Pal et al. demonstrated that neural networks trained in password similarity models were capable of compromising 16% of accounts tested in fewer than 1,000 attempts, further illustrating the inherent risk of predictable password patterns [14].

### 3.3. API Security Failures and Lack of Rate Limiting

The 23andMe breach was further exacerbated by security flaws in its authentication infrastructure. According to an analysis by Akto, a company specializing in API security, 23andMe's log-in API did not implement rate limiting, allowing attackers to make an unrestricted number of log-in attempts without triggering security mechanisms [15]. This oversight amplified the effectiveness of the credential-stuffing attack, making it easier for adversaries to test stolen credentials at scale systematically.

Although 23andMe officially confirmed that credential stuffing was the primary cause of the breach [16], the lack of robust API security measures contributed significantly to the scope of the compromise. Similar vulnerabilities have led to large-scale data breaches in other platforms, including the LinkedIn API scraping incident [17] and the Parler data breach, where inadequate authentication controls exposed sensitive user data [18].

The 23andMe breach underscores the critical importance of enforcing strict API security policies, including rate limiting, multi-factor authentication (MFA), and continuous monitoring for anomalous login behavior. These measures are essential to mitigating credential-surfing attacks and protecting user accounts from unauthorized access.

The 23andMe data breach had far-reaching consequences, exposing sensitive user information and raising significant cybersecurity, privacy, and ethical concerns. While only 14,000 user accounts were directly compromised, the breach's impact extended far beyond this initial figure due to the interconnected nature of the platform's data-sharing features. This section examines the extent of the data exposure, potential risks associated with the breach, and the broader societal implications, particularly concerning vulnerable demographic groups.

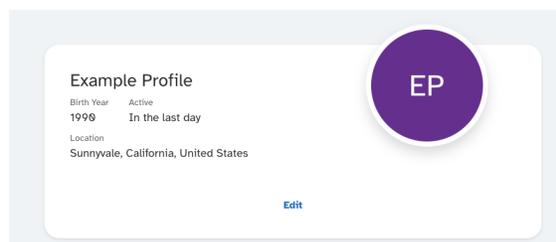

Figure 3: In application view of a sample profile for a related individual.

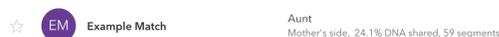

Figure 4: Another sample GUI element, showing some of the information shared even in non-confirmed relatives or matches.

### 3.4. Scale of the Breach and Data Exposure

According to 23andMe's official statements and blog updates, the attackers gained access to approximately 14,000 user accounts, representing less than 0.1% of the company's 14 million customers [16]. However, the breach extended well beyond these initially compromised accounts, as attackers leveraged access to these accounts to extract additional data from the DNA Relatives and Family Tree features.

The DNA Relatives feature allows users to opt-in for genetic matching, enabling them to connect with potential relatives based on shared DNA. This functionality inherently exposes sensitive information to connected users, creating a ripple effect when an account is breached. The data shared within this feature [19] includes:

1) User's DNA Relatives display name
2) Recent login activity
3) Relationship labels (masculine, feminine, neutral)
4) Predicted genetic relationship and percentage of shared DNA
5) Ancestry reports and matching DNA segments (optional)
6) User's location (optional)
7) Ancestor birth locations and family names (optional)
8) Profile picture (optional)
9) Birth year (optional)
10) Link to Family Tree (optional)
11) Personal notes in the "Introduce Yourself!" section (optional)

Due to these extensive data-sharing features, the 14,000 compromised accounts enabled attackers to access the profiles of approximately 5.5 million additional users. Moreover, 1.4 million Family Tree profiles containing detailed genealogical data were also compromised. This amplification of the breach highlights the significant privacy risks associated with genetic data-sharing platforms, where even a small-scale credential compromise can result in widespread data exposure.

### 3.5. Potential Data Extortion and Cybercrime Risks

The 23andMe breach raises concerns about how stolen genetic data might be exploited by cybercriminals. While financial data breaches often lead to fraudulent transactions, the unauthorized access to genetic data presents more complex and long-term threats.

One potential risk is data extortion, where attackers threaten to release sensitive genetic information unless victims pay a ransom. Similar tactics have been observed in high-profile cyber extortion cases, such as the DC Metropolitan Police ransomware attack by Babuk [20] and the Colonial Pipeline ransomware incident [21], where attackers sought financial gain or geopolitical leverage by exploiting stolen data. Given that genetic information is deeply personal and immutable, the consequences of such extortion schemes could be psychologically and socially damaging to victims.

Additionally, there is the risk of genetic discrimination, where employers, insurers, or other entities could misuse leaked genetic data to make biased decisions about individuals' health risks or ancestry. While laws such as the Genetic Information Nondiscrimination Act (GINA) in the U.S. provide some protection, the global availability of this data on illicit marketplaces could bypass legal safeguards.

#### 3.5.1. Ethical and Demographic Implications.
Beyond the technical and financial risks, the breach has also raised ethical concerns, particularly regarding its impact on specific demographic groups. While 23andMe has not publicly disclosed the full demographic breakdown of affected users, reports indicate that the attack disproportionately impacted individuals of Jewish and Chinese descent [22].

It remains unclear whether the attackers deliberately targeted these populations or if their data was simply more exposed due to network connections within the platform's DNA Relatives feature. However, regardless of intent, the breach highlighted systemic weaknesses in protecting sensitive demographic and ancestral data. Given the historical misuse of genetic information for racial profiling and discrimination, this incident underscores the urgent need for stronger privacy controls and data security measures in genetic testing services.

The 23andMe breach is a cautionary case study in how seemingly minor security flaws can lead to large-scale data exposure in interconnected platforms. While the initial 14,000 compromised accounts may seem limited in scope, the breach ultimately affected millions of users, demonstrating the far-reaching implications of genetic data-sharing systems. Moreover, the potential for data extortion, misuse, and discrimination emphasizes the need for stricter security measures, regulatory oversight, and user awareness. Moving forward, genetic testing companies must prioritize enhanced authentication mechanisms, stricter access controls, and improved transparency in data-sharing practices to prevent future incidents of this magnitude.

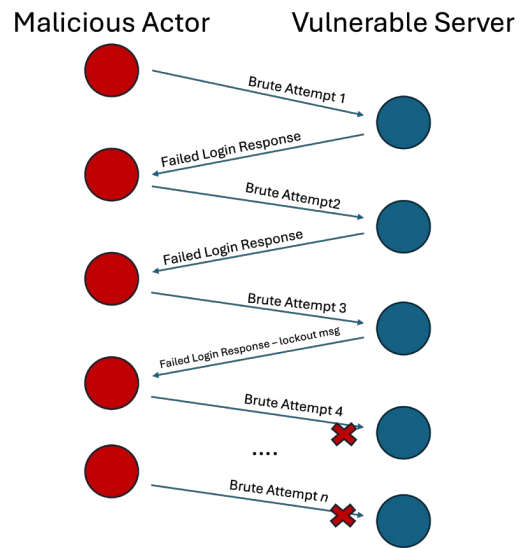

Figure 5: A basic example of rate limiting and how it functions to shut down repeated login attempts.

## 4. Defense Solutions

This section explores potential defense mechanisms against credential stuffing and brute-force attacks, as well as the security measures implemented by 23andMe in response to the breach. Modern cyberattacks often leverage multiple attack vectors to maximize their impact, as seen in the Astoria data breach [23] and the Accellion File Transfer Appliance ransomware attack [24], where attackers exploited credential theft, API vulnerabilities, and extortion tactics. Implementing robust security measures is crucial to mitigating similar threats.

### 4.1. Rate Limiting

One of the most effective countermeasures against brute-force attacks is *rate limiting*, which restricts the number of repeated login attempts within a given timeframe. Heiding *et al.* demonstrated that rate limiting significantly enhances security by mitigating brute-force attempts that systematically guess passwords [25]. While rate limiting does not entirely eliminate credential stuffing—especially when attackers use previously leaked credentials—it substantially increases the time required to carry out an attack, thereby reducing its feasibility. An example of rate limiting is illustrated in Figure 5.

The absence of rate limiting in 23andMe's authentication system represented a critical vulnerability. Tools such as Fail2Ban, a widely used rate-limiting software, are often recommended in security configurations for preventing automated login attempts. Heiding *et al.* conducted an experiment on honeypot systems and found that implementing Fail2Ban rules reduced SSH login attempts by 99.2% [25]. Specifically, the average number of attacks per IP address decreased from 101.1 to 4.6. While this figure slightly exceeds the Fail2Ban threshold of three attempts, minor delays in system responses occasionally permitted additional login attempts [25].

### 4.2. Password Complexity and Reuse Prevention

Brute-force and credential surfing attacks exploit human tendencies to create weak passwords or reuse credentials across multiple platforms. Weak password policies exacerbate security risks, as attackers can leverage large-scale data breaches to gain unauthorized access. Shay *et al.* studied various password policies and identified the *3class12* standard—requiring passwords to be at least 12 characters long and contain at least three character classes (letters, numbers, and special symbols)—as one of the most effective in balancing security and user recall [26]. This standard aligns with real-world security practices and helps mitigate credential-stuffing attacks.

Beyond enforcing stronger passwords, organizations must implement measures to prevent users from reusing compromised credentials. Thomas *et al.* developed a cloud-based security service integrated with a Chrome extension that alerts users when their passwords appear in known breaches [27]. In their study, 26% of users who received such alerts proactively changed their passwords. Table 2 presents a summary of their findings.

Although 26% may seem like a modest percentage, it translates to approximately 83,000 password resets initiated due to security alerts. The study also revealed that passwords associated with high-risk accounts, such as email, government, and shopping services, were more frequently changed than those linked to lower-risk categories like entertainment and adult websites [27]. These findings underscore the importance of proactive breach notification systems in improving password hygiene.

### 4.3. Passwordless Authentication and Multi-Factor Authentication

To further enhance security, 23andMe could adopt *passwordless authentication*, which allows users to access their accounts using alternative verification methods instead of traditional passwords [28]. This authentication model relies on biometric data (e.g., fingerprints or facial recognition), proximity devices (e.g., smart badges), or hardware tokens that generate time-sensitive codes. When integrated with Multi-Factor Authentication (MFA) and Single Sign-On (SSO) systems, passwordless authentication not only enhances security but also improves user experience and reduces IT overhead.

MFA is a critical defense against credential stuffing, requiring users to authenticate through multiple verification steps. Prior research has explored various MFA implementations, including password breach alerting and passwordless authentication mechanisms [29], [30].

### 4.4. Two-Step Verification

Following the breach, 23andMe implemented *two-step verification* as a mandatory security measure [16]. Two-step verification requires users to authenticate through two distinct factors: the first step involves entering traditional login credentials, while the second step requires a verification code sent via email, SMS, or an authentication app such as Microsoft Authenticator. This method significantly increases account security by introducing an additional barrier to unauthorized access.

Table 2: Results/findings from the usage of password vulnerability notifications. recreated from [27].

| Metric | Value |
|---|---|
| Extension Users | 667,716 |
| Logins Analyzed | 21,177,237 |
| Domains Covered | 746,853 |
| Breached Credentials Found | 316,531 |
| Warnings Ignored | 81,468 (26%) |
| Passwords Reset | 82,761 (26%) |

While two-step verification enhances security, it also presents potential availability challenges. As noted in [31], requiring a secondary authentication factor necessitates access to an external device, which may impact usability if the device is lost or unavailable. However, this issue can be mitigated through secure account recovery options.

Another concern, explored by Dmitrienko *et al.*, is that one-time passwords (OTPs) used in MFA systems may be intercepted by attackers [32]. Although OTP interception requires multiple layers of malicious activity, it remains a potential security risk. Nevertheless, MFA provides a substantial security improvement compared to password-only authentication.

### 4.5. Implementation of Security Measures at 23andMe

In response to the breach, 23andMe required all users to reset their passwords and implemented email-based two-step verification as a long-term security measure [33]. While these actions were taken post-incident, they represent a step in the right direction toward improving security practices. A more proactive approach—including the earlier implementation of rate limiting, passwordless authentication, and stronger password policies—could have mitigated the severity of the attack.

Moving forward, organizations handling sensitive user data must prioritize robust authentication protocols, continuous security monitoring, and user education on cybersecurity best practices. Strengthening the security culture within companies like 23andMe is essential to preventing similar incidents in the future.

## 5. Conclusion

The 23andMe data breach represents a significant cybersecurity failure, affecting millions of users and eroding trust in the company's ability to safeguard sensitive genetic and personal information. As 23andMe expanded its customer base, the attack by Golem not only compromised user data but also exposed vulnerabilities in the company's authentication and security measures. Despite a prompt response, the breach resulted in the unauthorized access of over 5.5 million customer records.

The nature of the compromised data underscores the severity of the breach. The exposed information included names, DNA relationships, ancestry reports, and identifying attributes such as relationship labels and ancestral family names. Such a data leak has profound implications, particularly for vulnerable populations. Furthermore, while 23andMe's official statements emphasized that only 14,000 accounts were directly compromised, the attack's cascading effect through the DNA Relatives feature significantly broadened its impact, leading to the exposure of millions of users' information.

The attack was executed through brute-force techniques and credential stuffing, exploiting weak authentication mechanisms and poor password hygiene. Golem's posts on BreachForums suggested that additional data may have been available for sale, further exacerbating concerns regarding the misuse of genetic and personal information.

The increasing prevalence of credential-stuffing attacks highlights the need for stronger security measures. Organizations and individuals must adopt proactive strategies to mitigate the risks associated with compromised credentials. Key defensive measures include rate limiting, the enforcement of strong and unique passwords, passwordless authentication, and multi-factor authentication (MFA). These techniques serve as fundamental safeguards against unauthorized access and data breaches.

Following the breach, 23andMe implemented mandatory password resets and enforced two-factor authentication for all users. However, the overall effectiveness of these measures ultimately depends on user compliance, particularly regarding password strength and reuse. Given that weak passwords and credential recycling remain common security risks, user education on password security is crucial in preventing future breaches.

This incident underscores the broader cybersecurity challenges faced by companies handling sensitive user data. The breach at 23andMe serves as a cautionary example of how inadequate authentication controls and insufficient security protocols can lead to large-scale data exposure. Moving forward, genetic testing companies and similar platforms must prioritize enhanced security policies, robust authentication mechanisms, and comprehensive user education to prevent similar breaches in the future.

# References

[1] "Dna genetic testing for health, ancestry and more - 23andme." https://www.23andme.com/.

[2] "What is 23andme?." https://www.yourdnaguide.com/what-is-23andme.

[3] Medusa, "23andme data leak: Brute force attack details and prevention."

[4] L. Franceschi-Bicchierai, "Hacker leaks millions more 23andme user records on cybercrime forum." https://techcrunch.com/2023/10/18/hacker-leaks-millions-more-23andme-user-records-on-cybercrime-forum/, 2023.

[5] L. Franceschi-Bicchierai, "23andme confirms hackers stole ancestry data on 6.9 million users." https://techcrunch.com/2023/12/04/23andme-confirms-hackers-stole-ancestry-data-on-6-9-million-users/.

[6] N. Mueller, "Credential stuffing." https://owasp.org/www-community/attacks/Credential_stuffing.

[7] "Dna relatives: The genetic relative basics." https://customercare.23andme.com/hc/en-us/articles/115004659068-DNA-Relatives-The-Genetic-Relative-Basics#:~:text=The%20DNA%20Relatives%20feature%20is,branch%20of%20your%20family%20tree.

[8] G. Cluley, "Millions of new 23andme genetic data profiles leak on cybercrime forum." https://www.bitdefender.com/blog/hotforsecurity/millions-of-new-23andme-genetic-data-profiles-leak-on-cybercrime-forum/, 2023.

[9] L. Bošnjak, J. Sreš, and B. Brumen, "Brute-force and dictionary attack on hashed real-world passwords," in *2018 41st international convention on information and communication technology, electronics and microelectronics (mipro)*, pp. 1161–1166, IEEE, 2018.

[10] D. Florêncio, C. Herley, and B. Coskun, "Do strong web passwords accomplish anything?," *HotSec*, vol. 7, no. 6, p. 159, 2007.

[11] Y. Li, H. Wang, and K. Sun, "A study of personal information in human-chosen passwords and its security implications," in *IEEE INFOCOM 2016 - The 35th Annual IEEE International Conference on Computer Communications*, pp. 1–9, 2016.

[12] M. H. N. Ba, J. Bennett, M. Gallagher, and S. Bhunia, "A case study of credential stuffing attack: Canva data breach," in *2021 International Conference on Computational Science and Computational Intelligence (CSCI)*, pp. 735–740, IEEE, 2021.

[13] C. Stejskal, A. Perminov, A. Lester, S. Bhunia, M. Salman, and P. A. Regis, "Analyzing the impact and implications of comb: A comprehensive study of 3 billion breached credentials," in *2024 IEEE/ACM 24th International Symposium on Cluster, Cloud and Internet Computing Workshops (CCGridW)*, 2024.

[14] B. Pal, T. Daniel, R. Chatterjee, and T. Ristenpart, "Beyond credential stuffing: Password similarity models using neural networks," in *2019 IEEE Symposium on Security and Privacy (SP)*, pp. 417–434, IEEE, 2019.

[15] "23andme data leak: Brute force attack details and prevention." https://www.akto.io/blog/23andme-data-breach-brute-force-attack-details-and-prevention.

[16] "Addressing data security concerns." https://blog.23andme.com/articles/addressing-data-security-concerns.

[17] B. Gibson, S. Townes, D. Lewis, and S. Bhunia, "Vulnerability in massive api scraping: 2021 linkedin data breach," in *2021 International Conference on Computational Science and Computational Intelligence (CSCI)*, pp. 777–782, IEEE, 2021.

[18] D. Redding, J. Ang, and S. Bhunia, "A case study of massive api scrapping: Parler data breach after the capitol riot," in *2022 7th International Conference on Smart and Sustainable Technologies (SpliTech)*, pp. 1–7, IEEE, 2022.

[19] "Dna relatives privacy and display settings." https://customercare.23andme.com/hc/en-us/articles/18262768896023-DNA-Relatives-Privacy-and-Display-Settings#:~:text=The%20DNA%20Relatives%20feature%20is,opt%20in%20to%20the%20feature.

[20] E. Caroscio, J. Paul, J. Murray, and S. Bhunia, "Analyzing the ransomware attack on dc metropolitan police department by babuk," in *IEEE International Systems Conference (SysCon)*, 2022.

[21] J. Beerman, D. Berent, Z. Falter, and S. Bhunia, "A review of colonial pipeline ransomware attack," in *2023 IEEE/ACM 23rd International Symposium on Cluster, Cloud and Internet Computing Workshops (CCGridW)*, pp. 8–15, IEEE, 2023.

[22] "23andme breach targeted jewish and chinese customers, lawsuit says." https://www.nytimes.com/2024/01/26/business/23andme-hack-data.html.

[23] J. Nadjar, Y. Liu, J. Salinas, and S. Bhunia, "A case study on the multi-vector data breach on astoria," in *2022 4th International Conference on Computer Communication and the Internet (ICCCI)*, pp. 51–57, IEEE, 2022.

[24] K. Kiesel, T. Deep, A. Flaherty, and S. Bhunia, "Analyzing multi-vector ransomware attack on accellion file transfer appliance server," in *2022 7th International Conference on Smart and Sustainable Technologies (SpliTech)*, pp. 1–6, IEEE, 2022.

[25] F. Heiding, R. Lagerström, A. Wallström, and M.-A. Omer, "Securing iot devices using geographic and continuous login blocking: A honeypot study," in *Proceedings of the 6th International Conference on Information Systems Security and Privacy 2020*, p. 424–431, INSTICC, 2020. Duplicate in Scopus 2-s2.0-85176319032QC 20201019.

[26] R. Shay, S. Komanduri, A. L. Durity, P. S. Huh, M. L. Mazurek, S. M. Segreti, B. Ur, L. Bauer, N. Christin, and L. F. Cranor, "Designing password policies for strength and usability," *ACM Trans. Inf. Syst. Secur.*, vol. 18, may 2016.

[27] K. Thomas, J. Pullman, K. Yeo, A. Raghunathan, P. G. Kelley, L. Invernizzi, B. Benko, T. Pietraszek, S. Patel, D. Boneh, *et al.*, "Protecting accounts from credential stuffing with password breach alerting," in *28th USENIX Security Symposium (USENIX Security 19)*, pp. 1556–1571, 2019.

[28] "Passwordless authentication." https://www.cyberark.com/what-is/passwordless-authentication/.

[29] H. He, J. Self, K. French, S. Bhunia, M. Salman, and P. A. Regis, "Zerologon explored: In-depth analysis and mitigation strategies for microsoft's critical vulnerability," in *2024 IEEE/ACM 24th International Symposium on Cluster, Cloud and Internet Computing Workshops (CCGridW)*, 2024.

[30] A. Prentosito, M. Skoczen, L. Kahrs, and S. Bhunia, "Case study on a session hijacking attack: The 2021 cvs health data breach," in *International Conference on Mobile Web and Intelligent Information Systems*, pp. 93–105, Springer International Publishing Cham, 2022.

[31] A. Dmitrienko, C. Liebchen, C. Rossow, and A.-R. Sadeghi, "On the (in)security of mobile two-factor authentication," in *Financial Cryptography and Data Security* (N. Christin and R. Safavi-Naini, eds.), (Berlin, Heidelberg), pp. 365–383, Springer Berlin Heidelberg, 2014.

[32] A. Dmitrienko, C. Liebchen, C. Rossow, and A.-R. Sadeghi, "Security analysis of mobile two-factor authentication schemes.," *Intel Technology Journal*, vol. 18, no. 4, 2014.

[33] "2-step verification for your 23andme account." https://customercare.23andme.com/hc/en-us/articles/360034119874-2-Step-Verification-For-Your-23andMe-Account.